\documentclass[prb, fleqn,twocolumn, showpacs, showkeys]{revtex4}

\newcommand{\ccro}{CaCu$_3$Ru$_4$O$_{12}$}

\usepackage[dvips]{epsfig}

\begin{document}

\title{Intermediate--valence behavior of the transition--metal oxide CaCu$_3$Ru$_4$O$_{12}$}

\author{A. Krimmel, A. G\"unther, W. Kraetschmer, H. Dekinger, N. B\"uttgen, V. Eyert, and A. Loidl}
\affiliation{Center for Electronic Correlations and Magnetism,
University of Augsburg, 86135 Augsburg, Germany}
\email{Alexander.Krimmel@physik.uni-augsburg.de}

\author{D. V. Sheptyakov}
\affiliation{Laboratory for Neutron Scattering, ETH Z\"urich $\&$
PSI Villigen, CH-5232 Villigen, Switzerland}

\author{E.-W. Scheidt and W. Scherer}
\affiliation{CPM, Institute of Physics, University of Augsburg,
86135 Augsburg, Germany}

\date{\today}

\begin{abstract}
The transition--metal oxide CaCu$_3$Ru$_4$O$_{12}$ with
perovskite--type structure shows characteristic properties of an
intermediate--valence system. The temperature--dependent
susceptibility exhibits a broad maximum around $150 - 160$~K. At
this temperature, neutron powder diffraction reveals a small but
significant volume change whereby the crystal structure is
preserved. Moreover, the temperature--dependent resistivity changes
its slope. NMR Knight shift measurements of Ru reveal a cross--over
from high temperature paramagnetic behavior of localized moments to
itinerant band states at low temperatures. Additional
density--functional theory calculations can relate the structural
anomaly with the $d$--electron number. The different experimental
and calculational methods result in a mutually consistent
description of CaCu$_3$Ru$_4$O$_{12}$ as an intermediate--valent
system in the classical sense of having low--energy charge
fluctuations.
\end{abstract}

\maketitle

$A$-site ordered perovskite oxides of stoichiometry
AA$^{\prime}_3$B$_4$O$_{12}$ show a variety of fascinating physical
properties. CaCu$_3$Ti$_4$O$_{12}$ has been extensively studied with
respect to colossal dielectric properties
\cite{Homes01,Lunkenheimer04} and CaCu$_3$Mn$_4$O$_{12}$ is a
ferromagnet with a record ordering temperature of $T_C=360$~K and
large magneto-resistance.\cite{Zeng99,Weht02} On the other hand, the
ruthenates ACu$_3$Ru$_4$O$_{12}$ (A=Na, Ca, La) are metallic Pauli
paramagnets \cite{Labeau80} displaying valence degeneracy.
\cite{Subramanian02} Moreover, they show enhanced Sommerfeld
coefficients indicative of strong electronic
correlations.\cite{Ramirez04,Kobayashi04,Krimmel08,Tanaka08}
Recently, CaCu$_3$Ru$_4$O$_{12}$ was shown to exhibit
heavy--fermion--like properties.\cite{Kobayashi04}  In analogy to
CeSn$_3$, a broad maximum in the susceptibility was taken as a
signature of the Kondo temperature of about 200 K and a moderately
enhanced Sommerfeld coefficient of $\gamma=85$~mJ/(f.u. mol K$^2$)
was observed.\cite{Kobayashi04} Conventional $f$-electron based
heavy--fermion states are formed via hybridization between two
different electronic subsystems, namely localized $f$--electrons and
band states of conduction electrons.\cite{Grewe91} This results in a
highly enhanced density of states (DOS) at the Fermi level $E_F$
(Abrikosov-Suhl resonance) and a concomitantly quasiparticle
renormalization described by an enhanced effective electronic mass
$m^*/m_e = 10^2 - 10^3$. However, it has been demonstrated that
heavy--fermion behavior should also occur in transition--metal
oxides upon approaching a metal--insulator transition.\cite{KvN03}

On the basis of x-ray spectroscopic measurements a direct analogy of
CaCu$_3$Ru$_4$O$_{12}$ to conventional $f$-electron heavy fermions
was reported, based on localized magnetic moments of the Cu$^{2+}$
ions and itinerant $d$-electrons originating from strong Ru-O
hybridization.\cite{Tran06} Localized and itinerant electrons are
supposed to be coupled antiferromagnetically by the Kondo mechanism.
\cite{Tran06} Recently, heat capacity and NMR/NQR measurements
revealed non--Fermi--liquid properties of \ccro at very low
temperatures.\cite{Krimmel08} Moreover, a comprehensive study of
ACu$_3$Ru$_4$O$_{12}$ (A=Na, Na$_{0.5}$Ca$_{0.5}$, Ca,
Ca$_{0.5}$La$_{0.5}$, La) (Ref. \onlinecite{Tanaka08}) established
large Sommerfeld coefficients $\gamma=75 - 136$~mJ/(f.u. mol K$^2$)
for all investigated compounds, whereas a broad maximum in the
susceptibility was observed only for \ccro. Evidently, \ccro
represents a particular case among electronically correlated
ruthenate compounds. Most recently, Kato {\it et al.} studied \ccro
by NMR experiments.\cite{Kato09} Following Ref.~\onlinecite{Kato09},
the temperature dependent relaxation rate excludes a localized
magnetic moment at the Cu site. Here we provide experimental
evidence that \ccro is an outstanding example of a transition--metal
oxide that can be classified as an intermediate valent (IV) system
in the classical sense of having low-energy charge fluctuations.

\begin{figure}[hbt] \centering
\epsfig{file=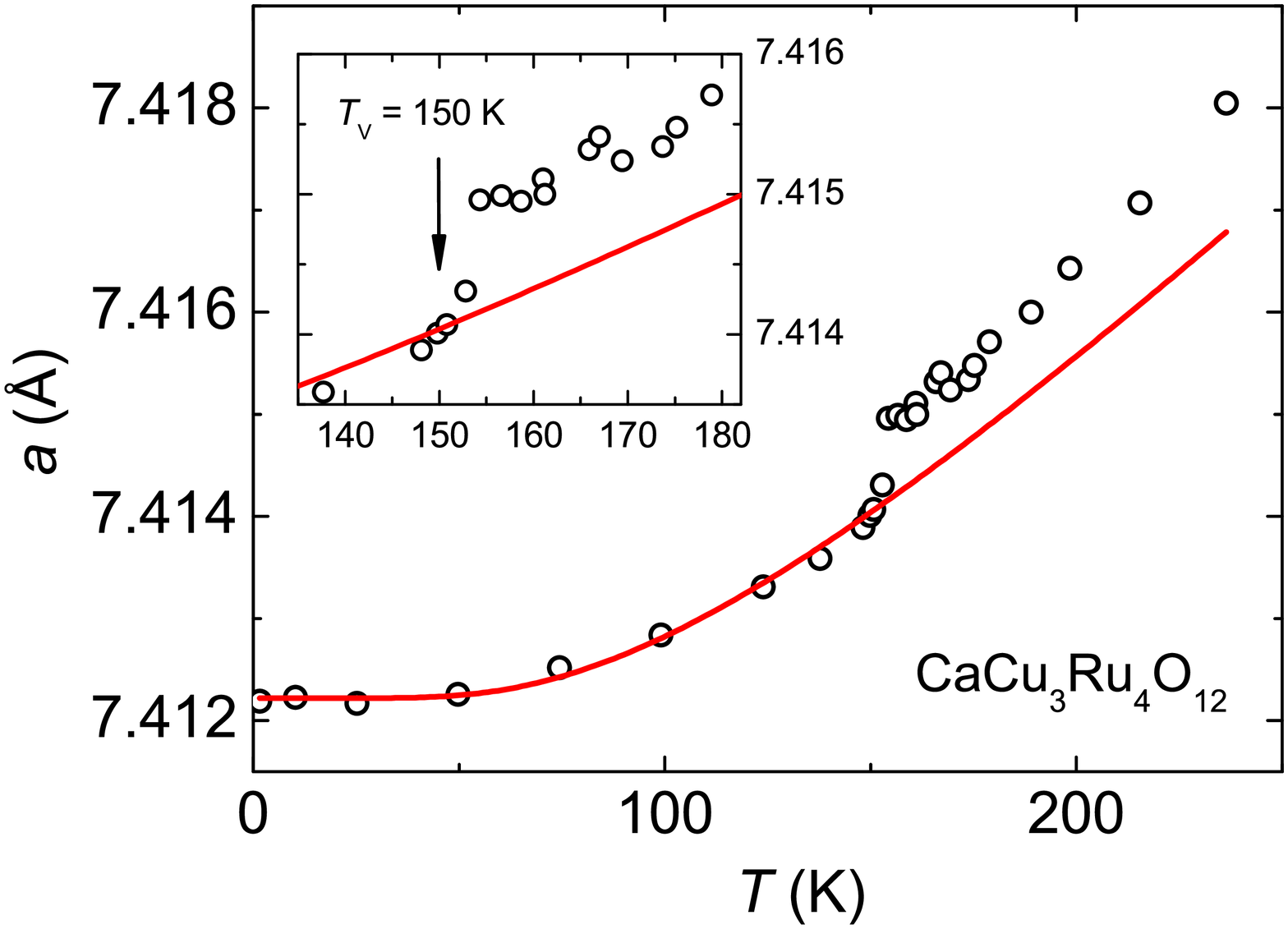,angle=0,width=0.46\textwidth} \caption{
Temperature dependence of the lattice constant $a$ of
CaCu$_3$Ru$_4$O$_{12}$ as resulting from the Rietveld refinements of
the neutron powder--diffraction measurements. The solid line is a
fit according to an usual anharmonic behavior described in the text.
}\label{hrpt}
\end{figure}

Polycrystalline samples of CaCu$_3$Ru$_4$O$_{12}$ were synthesized
by a solid state reaction as described previously.\cite{Krimmel08}
The samples were characterized by x-ray powder diffraction and
revealed single phase material without any indications of spurious
phases. Neutron powder diffraction measurements were performed on
the high resolution powder diffractometer (HRPT) for thermal
neutrons at the Paul Scherrer Institut, Switzerland. The diffraction
patterns were refined by standard Rietveld analysis employing the
FULLPROF program suite \cite{Rodriguez93}. The crystal structure of
CaCu$_3$Ru$_4$O$_{12}$ can be considered as a $2 \times 2 \times 2$
superstructure of the parent perovskite structure ABO$_3$. It is
described by cubic symmetry with space group $Im$\={3} and atomic
positions of Ca at (0, 0, 0), Cu at ($\frac{1}{2}, 0, 0$), Ru at
($\frac{1}{4}, \frac{1}{4}, \frac{1}{4}$) and O at ($x, y, 0$). At
$T=1.6$~K, refined structural parameters are the lattice constant
$a=7.41221(5)$~\AA ~and the oxygen positional parameters
$x=0.17478(13)$ and $y=0.30746(13)$. These values comply with those
reported in the literature \cite{Ebbinghaus02,Labeau80}. The
temperature dependent lattice constant as resulting from the
refinements is shown in Fig.~\ref{hrpt}. In this figure, the error
bars are smaller than the symbol size. A small ($ \approx 10^{-3}$)
but significant and sharp anomaly is evident at $T_V=150$~K. At low
temperatures $T\le150$~K, the lattice constant $a(T)$ was fitted
assuming an ordinary anharmonic behavior according to
$a(T)=a_0+\Delta a/(1-$exp$(-\theta_D/T))$ with a Debye temperature
of $\Theta_D=307$~K. The result is shown as a solid line in
figure~\ref{hrpt}. This fit nicely illustrates the abrupt expansion
of the unit--cell volume. Cycling the temperature revealed full
reversibility without hysteresis. This is in contrast to the
behavior of the homologue compound PrCu$_3$Ru$_4$O$_{12}$ which does
not show such a structural anomaly (not shown). The Rietveld
analysis shows that the crystal structure (lattice symmetry and
atomic positions) remains unchanged. Within the experimental
accuracy, the oxygen positional parameters of \ccro are constant.
The metal--oxygen distances show a smooth temperature dependence
whereas all metal--metal distances reveal an anomalous expansion
close to 150 K. A structural anomaly in form of a sudden lattice
expansion, whereby the crystal structure is fully preserved, clearly
points to an electronic origin in form of valence fluctuations. The
change in the unit--cell volume is consistent with a transition of
localized 4$d$--electrons at high temperatures ($T>T_V$) to
itinerant band states at low temperatures ($T<T_V$). Moreover, the
temperature of the structural transition of \ccro coincides
reasonably well with the temperature of the broad maximum of the
magnetic susceptibility, thus confirming an electronic origin.

\begin{figure}[hbt] \centering
\epsfig{file=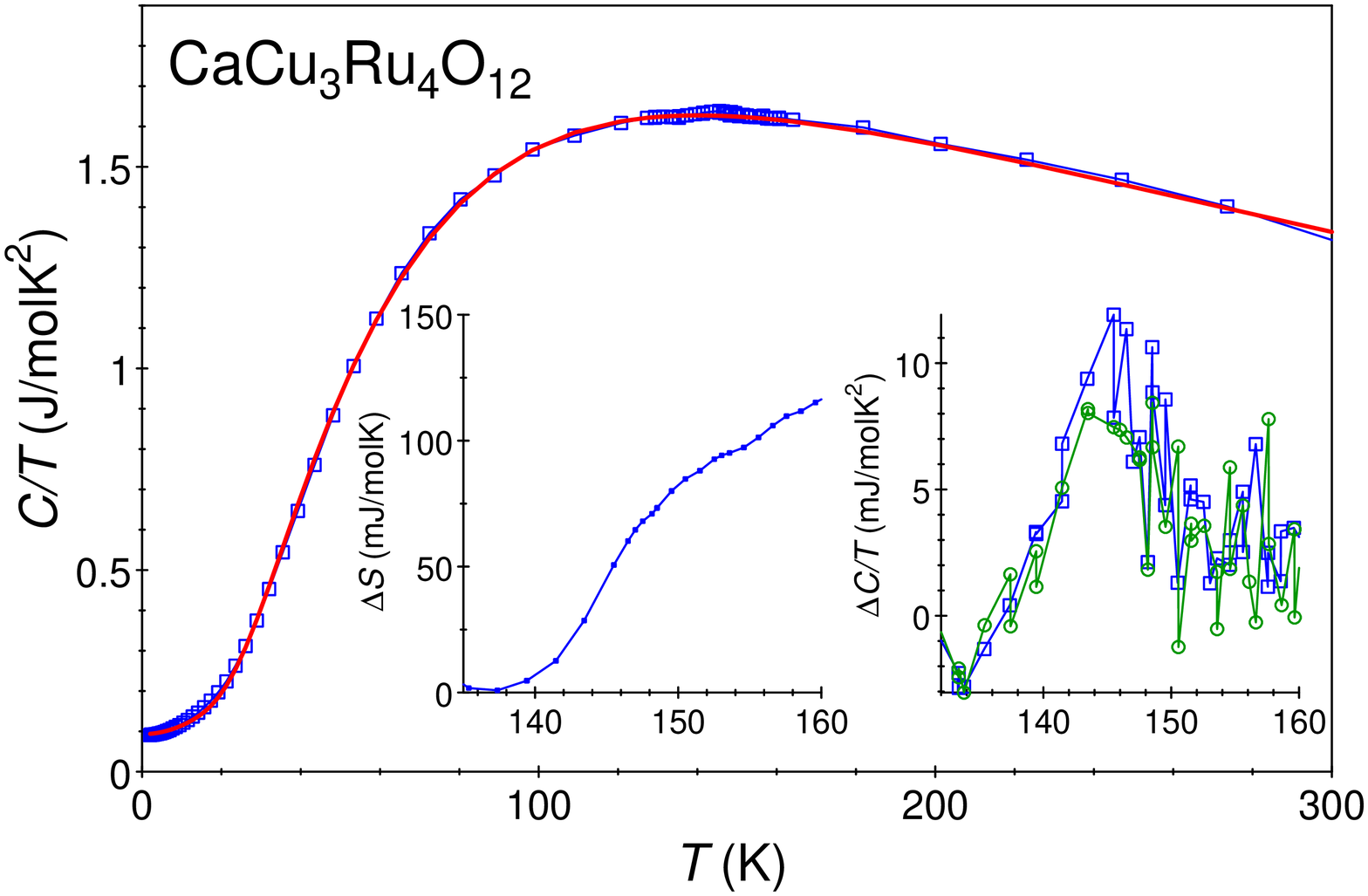,angle=0,width=0.46\textwidth} \caption{ Heat
capacity $C/T$ versus $T$ of CaCu$_3$Ru$_4$O$_{12}$. A small but
significant peak is observed in the temperature range $135 < T <
160$~K. To account for the phonon contributions, the data were
fitted by a Debye term and four Einstein modes (solid red line). The
right hand inset shows the heat capacity anomaly on an expanded
scale for heating (open blue squares) and cooling (open green
circles) evidencing the absence of hysteresis. The left hand inset
shows the associated entropy change of about 70 mJ/(f.u. mol
K).}\label{cp}
\end{figure}

The heat capacity $C(T)/T$ versus $T$ of CaCu$_3$Ru$_4$O$_{12}$ is
shown in Fig.~\ref{cp}. It exhibits a conventional behavior of
metallic Fermi liquids for $T>2$~K according to $C(T)=\gamma T +
\beta T^3$. Non--Fermi--liquid properties were observed
\cite{Krimmel08} for $T<2$~K. The phononic contribution results in a
Debye temperature of $\Theta_D=451$~K which is considerably enhanced
when compared to that determined from the low--temperature thermal
expansion. The electronic part is described by a Sommerfeld
coefficient of $\gamma=92 \pm 2$~mJ/(f.u. mol K$^2$). The right side
inset shows the specific heat of \ccro on an expanded scale in a
restricted temperature range $135 \le T \le 160$~K around the
anomalous volume transition. A small but significant maximum in
$C/T~(T)$ is seen. Heating and cooling runs (corresponding to open
blue squares and open green circles respectively) demonstrate a
complete reversibility of the volume transition without any
indications of hysteresis, as observed in the neutron
powder--diffraction measurements. This points towards a
second--order phase transition. The associated entropy is shown in
the left side inset and amounts to approximately 70 mJ/(f.u. mol K).

\begin{figure}[hbt] \centering
\epsfig{file=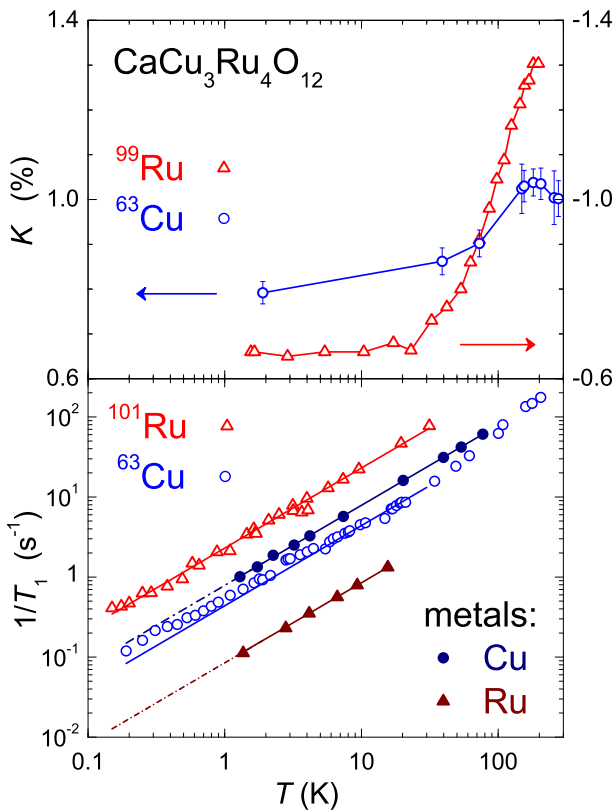,angle=0,width=0.45\textwidth} \caption{The
top frame shows the Knight shift of Cu (blue circles, left scale)
and Ru (red triangles, right scale), respectively. Please note the
different sign of the two scales. The bottom frame shows the
spin--lattice relaxation rate $1/T_1$ of \ccro for $^{63}$Cu and
$^{101}$Ru, respectively. The full lines represent fits to the data
according to a Korringa law. For comparison, the relaxation rates of
pure Cu and Ru metal are also shown.}\label{NMR}
\end{figure}

Electronic structure calculations \cite{Schwingenschlogl03} gave
evidence for strong covalent bonding between metal $d$ and oxygen
$p$ electrons. While the Ru-O bonds determine the size of the
RuO$_6$ octahedra, the Cu-O bonds dominate the octahedral tilting.
The partial DOS of Cu showed sharp peaks in the region $-2> E >
-3$~eV below the Fermi energy and some residual states above $E_F$
due to hybridization effects. This implies a rather closed 3$d$
shell and is consistent with the absence of a localized magnetic
moment at the Cu site in agreement with experiments.\cite{Kato09} In
the region around the Fermi energy, the electronic states derive
mainly from broader Ru 4$d$ bands although covalent bonding leads to
finite oxygen 2$p$ contributions at
$E_F$(Ref.\onlinecite{Schwingenschlogl03}). Therefore, a significant
contribution of the Kondo effect at the Cu site for the mass
renormalization can be excluded due to the absence of a magnetic
moment. The DOS at $E_F$ is dominated by Ru, whereas the Cu
contribution is negligible.\cite{Schwingenschlogl03} This is
experimentally confirmed by NMR (Knight shift) and NQR
(spin--lattice relaxation rate $1/T_1$) measurements on $^{63}$Cu,
$^{99}$Ru ,and $^{101}$Ru respectively. The bottom frame of
Fig.~\ref{NMR} shows the temperature--dependent spin--lattice
relaxation rate $1/T_1$ which reveals an enhanced Korringa behavior
for $^{101}$Ru [as compared to pure Ru metal,
(Ref.~\onlinecite{Mukuda99})] reflecting a high DOS at the Fermi
level, whereas $1/T_1$ of $^{63}$Cu is close but slightly below that
of pure Cu metal (Ref.~\onlinecite{Hanabusa72}).

The top frame of Fig.~\ref{NMR} shows the Knight shift $K$ of Cu
(left scale) and Ru (right scale), respectively. The Knight shift of
Cu is positive despite the fact that the core electron polarization
provides negative values. This excludes a significant contribution
of $d$--electrons to the Knight shift at the Cu site. The
temperature dependence of $^{Cu}K$ is weak and follows the magnetic
susceptibility. Taking into account the absence of a localized
magnetic moment atthe Cu site, we conclude that $^{Cu}K$ reflects
the temperature dependence of a transferred field. On the other
hand, the Knight shift at the Ru site reveals negative values due to
the Ru core--polarization effects. This points to a dominant
contribution of 4$d$--electrons of Ru ions to the magnetic behavior
of \ccro. Moreover, $^{Ru}K$ exhibits a more pronounced temperature
dependence evidencing a stronger moment compensation upon cooling.
Towards the lowest temperatures, the Knight shift at the Ru site
becomes temperature independent. This behavior hallmarks the
cross--over from paramagnetism of localized electrons to Pauli
paramagnetism of itinerant band states.

We further note that the temperature dependent resistivity changes
slope around $150 - 160$~K with a steeper decline in the low
temperature region.\cite{Krimmel08} Such a behavior is expected for
a mixed--valence transition when a fraction of originally localized
4$d$--electrons becomes (partly) itinerant.

New {\it ab initio} electronic structure calculations of \ccro were
performed with the structural input parameters obtained by the
refinements of the neutron powder--diffraction measurements. The
calculations are based on density--functional theory (DFT) and the
local--density approximation (LDA). They were performed using
scalar-relativistic implementation of the augmented spherical wave
(ASW) method.\cite{Eyert00,Eyert07} In the present work, a
full--potential version of the ASW method was employed in which the
electron density and related quantities are given by spherical
harmonics expansions inside the muffin--tin spheres. In the
remaining interstitial region, a representation in terms of
atom--centered Hankel functions is used. The calculated partial DOS
do not show any significant changes as compared to the previous
study \cite{Schwingenschlogl03}. The calculated $d$--electron number
of Ru has an almost half--integral value of 5.5 which reflects a
strongly covalent Ru-O bond and thus intermediate valency of the Ru
ions. Contrary, the nearly integral $d$-electron value of Cu
indicates a stronger ionic character of the Cu-O bond. A qualitative
difference between a low temperature ($T \le 125$~K) region with
almost constant $d$--electron number and a high temperature ($T \ge
150$~K) regime with a roughly linear increase of $n_d$ can be
distinguished.

It is evident that the transition between the high-$T$ and low-$T$
behavior of CaCu$_3$Ru$_4$O$_{12}$ involves only small changes in
the electronic structure. This is reflected in the small value of
10$^{-3}$ of the abrupt volume change at $T_V = 150$~K (see
Fig.~\ref{hrpt}). Moreover, the calculated variations of the
$d$--electron occupancies of about 10$^{-3}$ are two orders of
magnitude smaller than in traditional $f$--electron--based
intermediate--valence systems. The relaxation rate and the Knight
shift in recent NMR measurements \cite{Kato09} was explained in
terms of a two--band model for the markedly different high-$T$ state
above the susceptibility maximum around 160 K and the low-$T$
behavior below about 20 K. A strong hybridization corresponding to
increasing correlations between Cu and Ru electrons would enable an
effective charge transfer between these two bands.\cite{Kato09}
Charge order, charge localization, and valence transitions are often
observed in transition--metal oxides, mainly in manganites,
vanadates and ruthenates. As a prototypical example we discuss the
valence transition which has recently been reported in
LaCu$_3$Fe$_4$O$_{12}$ which is isostructural to
CaCu$_3$Ru$_4$O$_12$.\cite{Long09} On cooling, the Cu ions of this
compound undergo a valence transition from Cu$^{2+}$ to Cu$^{3+}$
accompanied by a metal to insulator transition and a striking volume
change. The corresponding volume change due to this integral change
in valence states is of the order of 1\% compared to a volume change
of 0.04\% in CaCu$_3$Ru$_4$O$_12$ where only a small fraction of $d$
electrons is involved. The susceptibility of the
AA$^{\prime}_3$B$_4$O$_{12}$ oxides toward charge disproportion with
associated structural and magnetic instabilities was also discussed
in the case of CaCu$_3$Fe$_4$O$_12$ recently.\cite{Hao09} However,
in all cases different but integral valence states have been
observed.

The $d$--electron correlations in transition--metal oxides are
traditionally taken into account by a Mott--Hubbard approach. The
relevant energy scale is determined by the on--site Coulomb
repulsion $U$ of the order of eV. Therefore, transition--metal
oxides usually exhibit high--energy electronic fluctuations. Many
years ago, Allen \cite{Allen85} discussed the possibility of valence
fluctuations in narrow--band oxides. The starting point was the
lattice Anderson Hamiltonian and its relation to a Hubbard band
description usually applied for transition--metal compounds. Valence
fluctuations become possible if the energy $E_g$ of the transition
$d^n \rightarrow d^{n+1}$ tends to zero, and it was suggested that
this situation may be realized near a metal--insulator or
metal--semimetal transition. The required large value of the
hybridization $V$ seems to favor sulfur (or even selenide) compounds
as compared to oxides. In the present case of \ccro, the strong
covalent Ru-O bond results from a strong hybridization between
ruthenium 4$d$ and oxygen 2$p$ electrons necessary for IV behavior.
The larger spatial extend of 4$d$ electron wave functions as
compared to the narrower 3$d$ electron bands seems to be a
prerequisite for a sufficiently strong hybridization to result in IV
properties but, on the other hand, weakens electronic correlations.
To look for further IV systems, one therefore should consider
4$d$--transition metal oxides with metallic conductivity and
enhanced Sommerfeld coefficients. These conditions seem to be
realized at best among the ruthenates.

In summary, the temperature dependent magnetic susceptibility, NMR
spin lattice relaxation rate $1/T_1$ and Knight shift measurements
of Ru and Cu, the anomalous volume transition at $T_V = 150$~K found
in neutron powder diffraction, which is confirmed by a corresponding
anomaly in the heat capacity, as well as electronic structure
calculations provide a mutually consistent description of
CaCu$_3$Ru$_4$O$_{12}$ as the first example of a transition--metal
oxide with IV properties in the classical sense of having
$d$--electron fluctuations. Other ruthenates seem to be the most
promising candidates to find further examples of this new class of
materials.

\acknowledgements This work was supported by the Deutsche
Forschungsgemeinschaft (DFG) via research unit 960 "Quantum Phase
Transitions", Sonderforschungsbereich 484 (Augsburg), DFG contract
SCHE487/7 and by the project COST P16 ECOM of the European Union.


\begin{references}

\bibitem{Homes01} C. C. Homes {\it et al.},
Science {\bf 293}, 673 (2001).

\bibitem{Lunkenheimer04} P. Lunkenheimer, R. Fichtl, S. G.
Ebbinghaus, and A. Loidl, Phys. Rev. B {\bf 70}, 172102 (2004) and
references therein.

\bibitem{Zeng99} Z. Zeng, M. Greenblatt, M. A. Subramanian, and M.
Croft, Phys. Rev. Lett. {\bf 82}, 3164 (1999).

\bibitem{Weht02} R. Weht and W. E. Pickett, Phys. Rev. B {\bf 65},
014415 (2001).

\bibitem{Labeau80} M. Labeau, B. Bochu, J. C. Joubert, and J.
Chenavas, J. Sol. State Chem. {\bf 33}, 257 (1980).

\bibitem{Subramanian02} M. A. Subramanian, and A. W. Sleight, Solid
State Sci. {\bf 4}, 347 (2002).

\bibitem{Ramirez04} A. P. Ramirez, G. Lawes, D. Li, and M. A.
Subramanian, Sol. State Comm. {\bf 131}, 251 (2004).

\bibitem{Kobayashi04} W. Kobayashi {\it et al.}, J. Phys. Soc. Jpn. {\bf 73}, 2373 (2004).

\bibitem{Krimmel08} A. Krimmel {\it et al.}, Phys. Rev. B {\bf 78},
165126 (2008).

\bibitem{Tanaka08} S. Tanaka {\it et al.}, J. Phys. Soc. Jpn. {\bf 78}, 024706
(2009).

\bibitem{Grewe91} For review, see for example N. Grewe and F.
Steglich {\it Handbook on the Physics and Chemistry of Rare
Earth}, Edts. K. A. Gschneidner Jr. and L. L. Eyring (Elsevier,
Amsterdam, 1991), Vol. {\bf 14}, p. 343.

\bibitem{KvN03}
H.-A. Krug von Nidda {\it et al.}, Eur. Phys. J. B {\bf 34}, 399
(2003).

\bibitem{Tran06} T. T. Tran {\it et al.}, Phys. Rev. B {\bf 73}, 193105 (2006).

\bibitem{Kato09} H. Kato {\it et al.}, J. Phys. Soc.
Jpn. {\bf 78}, 054707 (2009).

\bibitem{Rodriguez93} J. Rodriguez-Carvajal, Physica B {\bf 192},
55 (1993).

\bibitem{Ebbinghaus02} S. G. Ebbinghaus, A. Weidenkaff, and R. J.
Cava, J. Solid State Chem. {\bf 167}, 126 (2002).

\bibitem{Schwingenschlogl03} U. Schwingenschl\"ogl, V. Eyert, and
U. Eckern, Chem. Phys. Lett. {\bf 370}, 719 (2003).

\bibitem{Mukuda99} H. Mukuda et al., Phys. Rev. B {\bf 60}, 12279
(1999).

\bibitem{Hanabusa72} M. Hanabusa and T. Kushida, Phys. Rev. B {\bf 5}, 3751
(1972).

\bibitem{Eyert00} V. Eyert, Int. J. Quantum Chem. {\bf 77}, 1007
(2000).

\bibitem{Eyert07} V. Eyert, {\it The Augmented Spherical Wave Method -
A Comprehensive Treatment}, Lecture Notes Phys. {\bf 719}
(Springer, Berlin Heidelberg 2007).

\bibitem{Long09} Y. W. Long {\it et al.}, Nature (London) {\bf 458}, 60 (2009).

\bibitem{Hao09} X. Hao {\it et al.}, Phys.
Rev. B {\bf 79}, 113101 (2009).

\bibitem{Allen85} J. W. Allen, J. Mag. Mag. Mater. {\bf 47-48}, 168 (1985).




\end{references}
\end{document}